\documentclass[12pt,a4]{article}

\usepackage{color,tikz}

\usepackage{amsmath,amssymb,amsthm,amsfonts}
\usepackage{mathrsfs,graphicx,texdraw,hyperref}

\def\neprod{\setbox0=\hbox{$\nearrow$}
  \box0\kern-1.6em\prod} %.05em  %-1.15em
\def\swprod{\setbox0=\hbox{$\swarrow$}%
  \,\,\raise.03em\box0\kern-1.18em\prod} %.05em  %-1.15em
\def\seprod{\setbox0=\hbox{$\searrow$}%
  \,\,\raise.00em\box0\kern-2.0em\prod} %.05em  %-1.15em

\def\openone{\leavevmode\hbox{\small1\kern-3.3pt\normalsize1}}

\begin{document}

\begin{center}
{\LARGE \bf NLS-type equations from quadratic pencil of Lax operators: negative flows }

\bigskip

{
Rossen I. Ivanov \footnote{E-mail: {\tt rossen.ivanov@tudublin.ie}}
}

\medskip

\noindent
{\it School of Mathematical Sciences, Technological University Dublin, City Campus, \\  Grangegorman Lower, Dublin, D07 ADY7, Ireland}
\\[25pt]

\end{center}

{\small{\bf  Abstract.}
We formulate and study an integrable model of Nonlinear Schr\"odinger (NLS)-type through its Lax representation, where one of the Lax operators is quadratic and the other has a rational dependence on the spectral parameter. We discuss the associated spectral problem, the Riemann-Hilbert problem formulation, the conserved quantities, as well as a generalisation for symmetric spaces. In addition we explore the possibilities for modelling with higher order NLS (HNLS) integrable equations and in particular, the relevance of the proposed system.
\\[5pt]
{\bf Key words:} Bi-Hamiltonian integrable systems, Derivative nonlinear Schr\"odinger equation, Nonlocal integrable equations, Simple Lie algebra, Hermitian symmetric spaces
}

\section{Introduction}

It is a well known fact that a huge number of the physically important equations emerge as flows of the AKNS hierarchy \cite{AKNS}. The Lax operator of the AKNS hierarchy is $sl(2)$-valued, linear  with respect to the spectral parameter $\lambda,$ generalising the Zakharov-Shabat hierarchy Lax operator \cite{ZS1,ZS2}. The first, and best known example is of course the Nonlinear Schr\"odinger Equation (NLS) which is associated to the positive flow (the associated $M$ operator being a ``quadratic polynomial'' with respect to the spectral parameter). The gauge-equivalent hierarchies include the Heisenberg equation hierarchy. For more information we refer to the textbooks \cite{FaTa,ZMNP,GVY} and the references therein. AKNS hierarchy allows matrix generalisations, generalisations on simple Lie algebras and symmetric spaces, for example \cite{G86,I04,KuFo,GKV,GeGr,APT} just to mention a few.

There are other interesting and important integrable equations, arising from the ``negative'' flows of the hierarchy, that is, flows arising from $M$ operators with negative powers of $\lambda$ as well as flows with more complicated dependence on $\lambda,$ see for example \cite{Pav,AGZ,Iv20}. It is also known that the famous Camassa-Holm \cite{CH93}, (or CH) and Degasperis-Procesi \cite{D-P,D-H-H}, (or DP) equations could be considered as negative flows of generalised AKNS-type hierarchies with Lax operators respectively in $sl(2)$ and $sl(3),$ \cite{ILO,CIL,CI17}.

When the Lax operator is ``quadratic'' with respect to $\lambda$ it is sometimes called a {\it quadratic bundle} or {\it quadratic pencil}. Among the equations arising are the Derivative NLS (DNLS or DNLS I) \cite{KN, GIK, For} as well as DNLS~II \cite{ChLi} and DNLS~III {\cite{GI0, GI} equations, and more recently it was established that the negative flows lead to a non-evolutionary NLS-type equation \cite{Fo,LeFo} also known as the Fokas-Lenells  (FL) equation. These equations, (including the FL equation) could also be extended on Hermitian symmetric spaces \cite{GGI,V,GILF}.

The equations coming from the negative flows of the hierarchies are usually in a non-evolutionary form (CH, DP, FL etc.) In this work we explore a system, arising from a spectral problem, such that the $M$-operator has simple poles. We demonstrate that it leads to equation(s) also in a non-evolutionary form and generalises on Hermitian symmetric spaces as well.

We introduce the integrable model through its the zero curvature (Lax) representation. Then we discuss the spectral problem, conserved quantities and Hamiltonian formulation, as well as a generalisation for symmetric spaces. In the last section we explore the possibilities for modelling with higher order NLS (HNLS) integrable equations and discuss the relevance of the proposed system in the modelling context.

\section{Higher order NLS equation from a Lax operator with a quadratic dependence of the spectral parameter}

We start with a Lax pair, whose scattering operator $L$ is quadratic with respect to the spectral parameter $\lambda$:
\begin{equation}\label{eq:Lax}\begin{split}
 i \frac{\partial \psi}{ \partial x }& =  L\psi = \left(\left(\frac{1}{2}\lambda^2 +p(x,t)\right)\sigma_3+\lambda Q  \right) \psi(x,t,\lambda), \\
i \frac{\partial \psi}{ \partial t }& =  M\psi = \frac{1}{\lambda^2 - \zeta^2} \left(\left(\frac{1}{2}\lambda^2 +w(x,t)\right)\sigma_3+\lambda U  \right) \psi(x,t,\lambda).
\end{split}\end{equation}
In this Lax representation $p$ and $w$ are scalar functions, $\zeta$ is a complex constant, and
\begin{equation}
\sigma_3= \left(\begin{array}{cc} 1 & 0 \\ 0 & -1  \end{array}\right), \quad Q= \left(\begin{array}{cc} 0 & q \\ r & 0  \end{array}\right), \quad U= \left(\begin{array}{cc} 0 & u \\ v & 0  \end{array}\right)
\end{equation}
are $2 \times 2$ matrices.  The compatibility condition
\begin{equation}\label{eq:LenFo2}
iL_t - i M_x +[L,M]=0
\end{equation}
gives the following relations between the quantities (constraints)
\begin{equation}\label{eq:Q}\begin{aligned}
& p=-qr, \qquad w=-\zeta^2 \partial_x^{-1}(qr)_t,  \qquad u=(1-\partial_t)q, \qquad v=(1+\partial_t) r
\end{aligned}\end{equation}
and the following equations which use the above constraints:
\begin{equation}\label{eq:Q2}\begin{aligned}
& \zeta^2 iq_t +iq_x+q_{xt}+2qr(q-iq_t)-2  \zeta^2 q \partial_x^{-1}(qr)_t=0   , \\
 -&\zeta^2 ir_t -ir_x+r_{xt}+2qr(r+ir_t)-2  \zeta^2 r \partial_x^{-1}(qr)_t=0  .
\end{aligned}\end{equation}

The physical applications are usually related to equations, similar to NLS. In order to achieve such similarity and to obtain a NLS type equation of higher order we exchange the roles of $x$ and $t,$ hence
\begin{equation}\label{eq:Q2xt}\begin{aligned}
& iq_t+\zeta^2 iq_x+q_{xt}+2qr(q-iq_x)-2  \zeta^2 q \partial_t^{-1}(qr)_x=0   , \\
  -&ir_t-\zeta^2 ir_x+r_{xt}+2qr(r+ir_x)-2  \zeta^2 r \partial_t^{-1}(qr)_x=0  .
\end{aligned}\end{equation}
Furthermore, the following reductions in the sense of Mikhailov, \cite{Mi} are possible:

R1: $ r=\pm \bar{q} $ leading to
\begin{equation}\label{eq:Q3r1}
 iq_t +\zeta^2 iq_x+q_{xt}\pm 2|q|^2(q-iq_x)\mp 2  \zeta^2 q \partial_t^{-1}(|q|^2)_x=0 .
\end{equation}

The integration operator $ \partial_t^{-1}$ could be understood as $\int_{-\infty}^{t} dt',$ which leads to a hysteresis term in the equation. When $\zeta=0$ the equation resembles the Lenells-Fokas equation. We assume for simplicity that all functions are from the Schwartz class $\mathcal{S}(\mathbb{R})$ in $x$ for all values of $t.$

R2: This is a non-local reduction, $r(x,t)=\pm q(-x,-t),$  the reduction group could be extended to include reflections of coordinates, see for example \cite{V1}. The equation is (only the $-x$ and $-t$ arguments are explicit)
\begin{equation}\label{Q3r2}
iq_t+ \zeta^2 iq_x +q_{xt}\pm 2q q(-x,-t)(q-iq_x) \mp 2  \zeta^2 q \int_{-\infty}^{t}\big(q q(-x,-t)\big)_x dt'=0 .
\end{equation}
This is a nonlocal equation with hysteresis.

%\section{Spectral problem and inverse scattering}\label{sec:spec}
The solution techniques, such as inverse scattering, are based on the spectral theory of the Lax operator $L$, defined in \eqref{eq:Lax}. The inverse scattering theory allows a reformulation of the scattering problem in the form of a Riemann-Hilbert Problem (RHP). We are not going to discuss the RHP approach in details. The method is explained for example in \cite{GVY,ZMNP}. However we point out that an important role there plays the so called normalization of the RHP.

Let us rewrite the Lax operator $L$ in terms of the eigenfunction $\xi(x,t,\lambda)= \psi \exp( \frac{i}{2}\lambda^2 \sigma_3 x):$
\begin{equation}\label{L1}
i\xi_x= \frac{\lambda^2}{2}[\sigma_3, \xi]+(p\sigma_3 + \lambda Q )\xi.
\end{equation} %where
%$$Q= \left(\begin{array}{cc} 0 & q \\ r & 0  \end{array}\right). $$
If the canonical normalization $\lim_{|\lambda|\to \infty } \xi(x,\lambda)=\openone$ of the RHP is possible, then the series expansion as $|\lambda|\to \infty$ is
\begin{equation}\label{xi-ser}
\xi(x,\lambda)=\openone + \frac{\xi_1(x)}{\lambda}+ \frac{\xi_2(x)}{\lambda^{2}}+\ldots,
\end{equation}
Then, the substitution of \eqref{xi-ser} in \eqref{L1} allows to obtain identities for each power of $\lambda.$ In particular, for $\lambda^1$ we obtain
\begin{equation}
\frac{1}{2}[\sigma_{3}, \xi_{1}]+Q=0, \quad \text{giving} \quad \xi_{1}= \left(\begin{array}{cc} 0 & -q \\ r & 0  \end{array}\right),
\end{equation}
for $\lambda^0$ we obtain
\begin{equation}
\frac{1}{2}[\sigma_{3}, \xi_{2}]+p\sigma_3+Q\xi_1=0. \end{equation}
As far as $[\sigma_3, \xi_2]$ does not have a diagonal part, it follows that $p=-rq$ which is the same as \eqref{eq:Q},
in other words the RHP for the system under consideration has a canonical RHP normalization.
This is the exact same condition obtained in \cite{GI0}, with a slightly different, but equivallent definition of $L.$ The direct and the inverse scattering is explained in the same paper. Due to the canonical normalisation, the RHP method follows its standard routine. The dispersion law, of course is specific for the system under consideration.

The non-canonical normalisation appears in situations where $\lim_{|\lambda|\to \infty } \xi(x,\lambda)$ depends on $x.$  For the DNLS (Kaup-Newell) equation \cite{KN} and the Fokas-Lenells equation \cite{LeFo} for example, $p=0$ and the RHP normalisation is not canonical. Then a gauge transformation is usually employed to bring the RHP to another one with a canonical normalisation.

\section{Conserved quantities and Hamiltonians}\label{sec:ham}

In this section we discuss the conserved quantities of the equation. It is convenient to write it in an evolutionary form in terms of the variables $u=q-iq_x,$ $v=r+ir_x$
\begin{equation}\label{eq:Q2}\begin{aligned}
& iu_t +\zeta^2 iq_x +2qru +2q w=0   , \\
 -&iv_t -\zeta^2 ir_x +2qrv +2   r w =0  , \\
 &w_t=-\zeta^2(qr)_x.
\end{aligned}\end{equation}
This form is of interest for potential applications (when $q=\pm \bar{r}$). The conserved quantities are then generated by the $M$-operator (with $t $ and $x$ exchanged). Following the approach of Drinfel'd and Sokolov  \cite{DS},  $M$ can be diagonalized, that is,
\begin{equation}\label{}
 T( i\partial_x - M)T^{-1}=i\partial_x - \frac{1}{\lambda^2 - \zeta^2}\left (\frac{1}{2} \lambda^2  \sigma_3 +h_{-1}\lambda+ \sum_{k=0}^{\infty} h_k \lambda ^{-k} \right ),
\end{equation}
where $$T=\openone+ \sum_{k=0}^{\infty} T_k \lambda ^{-k},$$ and $h_k$ are diagonal matrices, representing the densities of the conserved quantities, while $T_k$ are off-diagonal matrices. As a result we obtain $h_{-1}=0,$ $h_{2k+1}=0,$
\begin{equation}\label{eq:h}\begin{aligned}
& h_0=\mathrm{diag}(ur+w,-vq-w),  \\
& h_2=\mathrm{diag}(u(1+i\partial_x)^{-1}(i\zeta^2 r_x-2wr-ur^2),v(1-i\partial_x)^{-1}(i\zeta^2 q_x+2wq+vq^2))
\end{aligned}\end{equation} giving rise to the conserved quantities
\begin{equation}\label{eq:Hh}\begin{aligned}
& H'_1=\int (ur+w)dx=\int [r(q-iq_x)+w]dx=\int [q(r+ir_x)+w]=\int (qv+w)dx,  \\
& H_2=-\int[\zeta^2 q i r_x -2wqr - r^2q(q-iq_x)]dx=\int [\zeta^2 irq_x +2wqr +q^2r(r+ir_x)]dx
\end{aligned}\end{equation} which play the role of the first two conserved quantities (Hamiltonians). The higher conserved quantities could only be given in terms of nonlocal densities. Moreover, from the third equation of the system it follows that $H^*=\int w dx $ is conserved automatically as well. This quantity is a Casimir, hence $H_1=\int ur dx=\int qv dx.$ The equations could be written in a Hamiltonian form as
 \begin{equation}\label{HamEq}
  \begin{pmatrix} iu_t \\
 iv_{t} \\
 iw_t \end{pmatrix} =  \mathcal{D}\begin{pmatrix}  \frac{\delta H_2}{\delta u}  \\
 \frac{\delta H_2}{\delta v} \\
 \frac{\delta H_2}{\delta w} \end{pmatrix} \quad \text{where} \quad
 \mathcal{D}= \begin{pmatrix}  0 &  -(1-i\partial_x) & 0 \\
 (1+i\partial_x) & 0 & 0 \\
 0 & 0 & \frac{i}{2} \zeta^2 \partial_x \end{pmatrix}
 \end{equation} is a Hamiltonian operator, which defines a Poisson Bracket.  The operator $\mathcal{D}$ is clearly Hamiltonian, as a direct sum of the Hamiltonian operator that appears in the Lenells-Fokas system, see \cite{LeFo}, and the Hamiltonian operator $\partial_x.$  The bi-Hamiltonian formulation necessitates the rigorous construction of the recursion operator of the system. The evolution under the Casimir is indeed trivial:
  \begin{equation}\label{HamEq}
  \begin{pmatrix} iu_{\tau} \\
 iv_{\tau} \\
 iw_{\tau} \end{pmatrix} = \mathcal{D} \begin{pmatrix}  \frac{\delta H^*}{\delta u}=0  \\
 \frac{\delta H^*}{\delta v} =0\\
 \frac{\delta H^*}{\delta w}=1 \end{pmatrix} = \begin{pmatrix} 0 \\
 0 \\
 0 \end{pmatrix} .
 \end{equation}

\section{Lax formulation for Hermitian symmetric spaces}\label{sec:sym}

The multi-component generalisations of NLS are heavily studied in soliton theory. A large class of matrix generalizations involve Lax pairs taking values in some simple Lie algebra. Furthermore, the simple Lie algebras admit splitting which ia associated to the structure of a Hermitian symmetric space. More details can be found in the classical book \cite{h} and in the seminal works of Athorne, Fordy and Kulish, \cite{AtFo,KuFo}. Integrable systems on symmetric spaces of finite dimensional Lie algebras have been studied considerably in the literature, see \cite{KuFo,AtFo,For,Basic,GeGr,GGK,GILF,ArHoIv}.

A simple Lie algebra $\frak{g}$ over the complex numbers admits the splitting \begin{align}\label{spl}
  \mathfrak {g}= \mathfrak{k}\oplus \mathfrak{m}\, ,
\end{align}
  where $\mathfrak{k}$ is a subalgebra of $\mathfrak{g},$ and $\mathfrak{m}$ is the complementary subspace of $\mathfrak{k}$ in $\mathfrak{g}.$
  In addition, \begin{align}
        [\mathfrak{k}, \mathfrak{k}] \subset \mathfrak{k}\, , \qquad [\mathfrak{k},\mathfrak{m} ] \subset  \mathfrak{m}\, , \qquad [\mathfrak{m},\mathfrak{m}] \subset \mathfrak{k}\, .
    \label{sym-relation}
\end{align}

 Denoting by $K$ and $G $ the Lie groups, associated to $\mathfrak{k} $ and $\mathfrak{g}$ correspondingly, the linear subspace $\mathfrak{m} $ is  identified with the tangent space of $G/K,$ which is used as a notation for the corresponding symmetric space.

The Hermitian symmetric spaces are a special class of symmetric spaces for which there is a special element $J\in \mathfrak{k} $ such that
\begin{equation}\label{Jel}
  \mathfrak{k}=\{ X \in \mathfrak{g } : \, \, [J,X]=0    \}, \quad [J, \mathfrak{k}]=0.
\end{equation}
It is clear then that the Cartan subalgebra $\mathfrak{h}\subset \mathfrak{k}\subset \mathfrak{g},$ and the element $J$ can be chosen from the Cartan
subalgebra, $J \in \mathfrak{h}.$ In other words, $J$ will be chosen to be diagonal. Furthermore, this element is highly degenerate, in a sense that  $\mathrm{ad}_J,$ which is an $n \times n$ matrix ($n=\mathrm{dim}(\mathfrak{g})$) has only three eigenvalues; $0$ and $\pm a,$ and the subspace $\mathfrak{m}$ can be split
further as
\begin{align*}
  \mathfrak{ m}= \mathfrak{m^+}\oplus \mathfrak{m^-}, \quad  \mathfrak{m^{\pm}}=\{X^{\pm}: \,\, [J, X^{\pm}]=\pm a X^{\pm} \}.
\end{align*}

The spectral problem is quadratic with respect to the spectral parameter $\lambda,$
\begin{equation}\label{eq:LA3}\begin{split}
 i \frac{\partial \psi}{ \partial x }& =  L\psi =(\lambda^2 J + \lambda Q + P) \psi(x,t,\lambda), \\
i \frac{\partial \psi}{ \partial t }& =  M\psi = \frac{1}{\lambda^2 - \zeta^2} \left(\lambda^2 J+\lambda U + W \right)  \psi(x,t,\lambda) ,
\end{split}\end{equation} where $L,M \in \mathfrak{g},$ $P,W \in \mathfrak{k}$ and $Q,U \in \mathfrak{m}.$

The matrix realisation of the symmetric spaces is with block matrices, such that the splitting \eqref{spl} is related to the matrix block structure for the corresponding symmetric space. For the {\bf A.III} symmetric space the block structure is

\begin{equation}\label{eq:LA3}\begin{split}
 i \frac{\partial \psi}{ \partial x }& =  L\psi =\left(\begin{array}{cc} \frac{1}{2}\lambda^2 \openone +p_1(x,t) & \lambda q \\ \lambda r & -\frac{1}{2}\lambda^2\openone +p_2(x,t)  \end{array}\right) \psi(x,t,\lambda), \\
i \frac{\partial \psi}{ \partial t }& =  M\psi = \frac{1}{\lambda^2 - \zeta^2} \left(\begin{array}{cc} \frac{1}{2}\lambda^2\openone +w_1(x,t) & \lambda u \\ \lambda v& -\frac{1}{2}\lambda^2\openone +w_2(x,t)  \end{array}\right)  \psi(x,t,\lambda) ,
\end{split}\end{equation}
where $q,r,u,v,p_1,p_2,w_1,w_2$ are matrices of corresponding dimensions, $\zeta$ is a constant. The compatibility condition
\begin{equation}\label{eq:LenFo2}
iL_t - i M_x +[L,M]=0
\end{equation}
leads to the following equations:
%\begin{equation}\label{eq:A3}\begin{aligned}
%& ip_{1,t}+qv-ur=0, \\
%&  ip_{2,t}+ru-vq=0,  \\
%& -\zeta^2 i p_{1,t}-iw_{1,x}+[p_1,w_1]=0   , \\
%& -\zeta^2 i p_{2,t}-iw_{2,x}+[p_2,w_2]=0  , \\
%& u=q-iq_t,\\
%& v=r+ir_t,\\
%& -\zeta^2iq_t-iu_x +p_1 u-w_1 q +q w_2 - u p_2 =0,\\
%& -\zeta^2ir_t-iv_x +rw_1 -vp_1 +p_2 v -w_2 r =0.\\
%\end{aligned}\end{equation}
%The equations yield
\begin{equation}\label{eq:A32}\begin{aligned}
& p_1=-qr, \qquad p_2= rq ,  \\
& iw_{1,x}-\zeta^2 i (qr)_t+[qr, w_1]=0, \\
& -iw_{2,x}-\zeta^2 i (rq)_t+[rq, w_2]=0, \\
& -\zeta^2 iq_t-i(q-iq_t)_x-qr(q-iq_t)-(q-iq_t)rq -w_1q+qw_2=0,\\
& -\zeta^2 ir_t-i(r+ir_t)_x+(r+ir_t)qr+rq(r+ir_t)+rw_1 -w_2r=0.
\end{aligned}\end{equation}
Again, as before, the change of the variables $x$ and $t$ gives
\begin{equation}\label{eq:A33}\begin{aligned}
& iw_{1,t}-\zeta^2 i (qr)_x+[qr, w_1]=0, \\
-& iw_{2,t}-\zeta^2 i (rq)_x+[rq, w_2]=0, \\
& iq_t +q_{xt}+\zeta^2 iq_x +qr(q-iq_x)+(q-iq_x)rq +w_1q-qw_2=0,\\
-& ir_t +r_{xt} -\zeta^2 ir_x+(r+ir_x)qr+rq(r+ir_x)+rw_1 -w_2r=0.
\end{aligned}\end{equation}

The reduction $r=q^{\dagger}$ leads to $w_1=w_1^{\dagger},$ $w_2=w_2^{\dagger},$ and the coupled system of equations

\begin{equation}\label{eq:A33}\begin{aligned}
& iw_{1,t}-\zeta^2 i (qq^{\dagger})_x+[qq^{\dagger}, w_1]=0, \\
-& iw_{2,t}-\zeta^2 i (q^{\dagger}q)_x+[q^{\dagger}q, w_2]=0, \\
& iq_t +q_{xt}+\zeta^2 iq_x +qq^{\dagger}(q-iq_x)+(q-iq_x)q^{\dagger}q +w_1q-qw_2=0.\\
\end{aligned}\end{equation}

This short example illustrates that the construction works for Hermitian symmetric spaces like in \cite{GILF}. A detailed study for specific symmetric spaces deserves a separate publication.

\section{Modelling with NLS-type equations}\label{sec:models}

\subsection{Higher order NLS (HNLS) equations, integrability and asymptotic expansions}

From modelling point of view, the generalised NLS or Higher order NLS (HNLS) equations with applications in nonlinear optics \cite{Ko,KoHa} as well in water waves \cite{GH} and plasma \cite{GT96} are

\begin{equation}\label{NLStype}
  iq_{T} +i c  q_X+\frac{1}{2}q_{XX} + |q|^2q+i\beta_1q_{XXX}+i\beta_2|q|^2q_X+i\beta_3 q (|q|^2)_X=0.
\end{equation}
where $c,$ $\beta_i,$ $i=1,2,3$ are, in general arbitrary real constants, depending on the physical parameters. Derivation of the HNLS in various situations could be found also in \cite{CEHS}.

The integrable cases correspond to the following ratios $(\beta_1: \beta_2:\beta_3):$  The DNLS I and II with $(0:1:1)$ and $(0:1:0)$ , see \cite{KN,ChLi}; Hirota \cite{Hi} with $(1:6:0)$ and Sasa-Satsuma \cite{SS91} with $(1:6:3).$

Nijhof and Roelofs \cite{NR92} using the prolongation method have proven that no other integrable cases in the form \eqref{NLStype} exist.
Therefore, in general, the exact values of the parameters is difficult to match. For practical applications, however, there is usually a small parameter, $\varepsilon,$ such that the quantities scale like $q\sim \varepsilon , $ i.e. $q\rightarrow \varepsilon q,$ $T=\varepsilon t,$ slow time, then $t=T/\varepsilon$ and $X=\varepsilon x$ - slow space variable, where $(x,t)$ are the original unscaled variables. Therefore, the physical models leading to HNLS involve perturbative expansions like
\begin{equation}\label{NLStype1}
  iq_{t} + ic q_x+\frac{\varepsilon}{2}q_{xx} + \varepsilon |q|^2q+i\varepsilon^2\left(\beta_1q_{xxx}+\beta_2|q|^2q_x+\beta_3 q (|q|^2)_x\right)=\mathcal{O}(\varepsilon^3).
\end{equation}
Thus we can extend the set of integrable models by considering those, which admit the scaling as in \eqref{NLStype1}, and we can use the triples $(\beta_1:\beta_2:\beta_3)$ for some systematic classification of the models. For example, the Fokas - Lenells equation \cite{LeFo}
\begin{equation}\label{LF}
  iq_t +icq_x-\varepsilon \nu q_{xt} + \varepsilon  \gamma q_{xx} + \varepsilon |q|^2 q + i\varepsilon^2 \nu |q|^2 q_x=0,
\end{equation}
where $c, \nu, \gamma$ are constants, can be transformed as follows. In the leading order, $q_t=-cq_x$ thus we have
\begin{equation}\label{LF0}
\begin{split}
  &iq_t +icq_x +\varepsilon (\gamma+\nu c) q_{xx} +  \varepsilon |q|^2 q =\mathcal{O}(\varepsilon^2),\\
  &q_t=-cq_x+i\varepsilon (\gamma+\nu c) q_{xx} + i \varepsilon |q|^2 q +\mathcal{O}(\varepsilon^2)
  \end{split}
\end{equation}
Next we substitute $q_t$ from \eqref{LF0} in the $q_{xt}$ term of \eqref{LF} to obtain
\begin{equation}\label{LF1}
 i q_t +icq_x+\varepsilon (c\nu+ \gamma) q_{xx} + \varepsilon |q|^2 q -i\varepsilon^2 \nu( c \nu+\gamma)q_{xxx}- i\varepsilon^2 \nu q( |q|^2 )_x=\mathcal{O}(\varepsilon ^3).
\end{equation}

In order to match \eqref{NLStype} we need to choose $c\nu+\gamma=1/2,$ then

\begin{equation}\label{LF2}
 i q_t +icq_x+ \frac{\varepsilon }{2} q_{xx} + \varepsilon |q|^2 q -i\varepsilon^2 \frac{\nu}{2}\big( q_{xxx} +2 q( |q|^2 )_x\big)=\mathcal{O}(\varepsilon ^3),
\end{equation}
hence the model is characterised by the ratio $(1:0:2).$

The asymptotic classification of the integrable models with the triples/ratios $(\beta_1:\beta_2:\beta_3)$ is useful for practical purposes but it is not unique since it depends on the chosen variables used in the model description. For example, the NLS model itself $(0:0:0)$ could be extended by the transformation $q= \mathfrak{a}+i\varepsilon \beta \mathfrak{a}_x$ where $\mathfrak{a}(x,t)$ is a new variable, and $\beta$ is a constant. Neglecting terms of order $\varepsilon^3$ this gives a model of type \eqref{NLStype1} for $\mathfrak{a}$ with $\beta_1=0,$  $\beta_2 = -2\beta$ and $\beta_3=2\beta .$

\subsection{The asymptotic expansion of \eqref{eq:Q3r1} and HNLS }

Let us write \eqref{eq:Q3r1} with the upper sign, introducing the scale factor:

\begin{equation}\label{Neq1}
iq_t+ \zeta^2 iq_x +\varepsilon q_{xt}+2\varepsilon |q|^2(q-i\varepsilon q_x)-2\varepsilon  \zeta^2 q \int_{-\infty}^{t}(|q|^2)_x dt'=0 .
\end{equation}

In the leading order $q_t=-\zeta^2 q_x$ thus we can relate the $x$ and $t$-derivatives in the nonlocal term to perform the integration and to obtain the next order approximation
\begin{equation}\label{new0}
\begin{split}
  q_t=-\zeta^2 q_x-i\varepsilon \zeta^2 q_{xx} +4 i \varepsilon |q|^2 q +\mathcal{O}(\varepsilon^2)
  \end{split}
\end{equation}
Next we substitute $q_t$ from \eqref{new0} in the $q_{xt}$ term of \eqref{Neq1} to obtain
\begin{equation}\label{newa}
  iq_t +i\zeta^2 q_x -\varepsilon \zeta^2 q_{xx} -i\varepsilon^2 \zeta^2q_{xxx}+4 i\varepsilon^2 ( q |q|^2 )_x +2\varepsilon |q|^2 (q-i\varepsilon q_x)
  -2 \varepsilon \zeta^2 q \partial_t ^{-1}(q_x \bar{q} +q \bar{q}_x) =\mathcal{O}(\varepsilon^3).
\end{equation}

Let us now consider the term $\partial_t ^{-1}(q_x \bar{q} +q \bar{q}_x) .$ From \eqref{new0} we express
\begin{equation}\label{new1}
\begin{split}
   q_x=-\frac{1}{\zeta^2} q_t-i\varepsilon q_{xx} +4 i \frac{\varepsilon}{\zeta^2} |q|^2 q +\mathcal{O}(\varepsilon^2)
  \end{split}
\end{equation} and substitute it in $\partial_t ^{-1}(q_x \bar{q} +q \bar{q}_x) :$
\begin{equation}\label{new2}
\begin{split}
 \partial_t ^{-1}(q_x \bar{q} +q \bar{q}_x) =-\frac{1}{\zeta^2} |q|^2 + i \varepsilon \partial_t^{-1} (q \bar{q}_x -q _x \bar{q}) _x +\mathcal{O}(\varepsilon^2)
  \end{split}
\end{equation}

Since in the leading order $q$ and all expressions containing it are functions of $x-\zeta^2 t$  we have
\begin{equation}\label{new3}
\begin{split}
 \partial_t ^{-1}(q_x \bar{q} +q \bar{q}_x) =-\frac{1}{\zeta^2} \left( |q|^2 + i \varepsilon  (q \bar{q}_x -q _x \bar{q})\right) +\mathcal{O}(\varepsilon^2),
  \end{split}
\end{equation} and therefore
\begin{equation}\label{newa}
  iq_t +i\zeta^2 q_x -\varepsilon \zeta^2 q_{xx} +4\varepsilon |q|^2 q  -i\varepsilon^2 \zeta^2q_{xxx}+6 i\varepsilon^2  q (|q|^2 )_x
  -2i\varepsilon^2 |q|^2 q_x
   =\mathcal{O}(\varepsilon^3).
\end{equation}

In order to match \eqref{NLStype}, after overall division by 4, the coefficient of $q_{xx}$ has to be $1/2,$
thus we need to choose $\zeta^2=-2.$ The first two linear terms can be transformed with a Galilean change and re-scaling of $t$ to a new time-like variable $\tau$ to $iq_{\tau},$ then

\begin{equation}\label{newf}
  iq_{\tau}+ \frac{\varepsilon }{2} q_{xx} + \varepsilon |q|^2 q +i\frac{\varepsilon^2 }{2}\big( q_{xxx} -|q|^2q_x+3 q( |q|^2 )_x\big)=\mathcal{O}(\varepsilon ^3),
\end{equation}
hence the model is characterised by the ratio $(1:-1:3).$

Following the same procedure, for the lower sign of equation \eqref{eq:Q3r1} we obtain the ratio $(1:-2:6)$ for $\zeta^2=2.$

%\section{Discussion}
\small
\section{Acknowledgements}
The author is partially supported by the Bulgarian National Science Fund, grant K$\Pi$ -06H42/2
from 27.11.2020. The author is thankful to two anonymous referees whose comments and suggestions have improved the quality of the manuscript.

%\small

\end{document}